# The Casimir Forces in a Single Conducting Cylindrical Cavity

**H. Razmi** (1) and **S. M. Shirazi** (2)

Department of Physics, the University of Qom, *3716146611*, Qom, I. R. Iran.

(1) razmi@qom.ac.ir & razmiha@hotmail.com (2) sms102@gmail.com

We want to study the Casimir effect for a single conducting microscopic cylindrical cavity. The mathematical technique is based on the Green function of the geometry of the inside of the cavity, and the integral regularization is based on the plasma frequency cutoff for real conductors. Using the symmetric electromagnetic energy-momentum tensor, in terms of four potential $A^{\mu}$, the total Casimir energy for the inside of the Cavity is calculated. Neglecting the contribution of the external (outside of the cavity) Casimir energy based on the reason recently presented in [1], the forces experienced by the lateral surface of the cavity and its circular bases are calculated. The resulting expressions show that the forces are repulsive. The numerical computation is done for the real problem of a cavity with a basis of a radius in the same order of its height at the scale of $100nm$ made of the best conducting materials already known.

**Keywords:** The Casimir Effect; Canonical Quantization; Green Function

**PACS:** 03.70.+k, 42.50.Pq

## Introduction

Quantum vacuum energy and the Casimir effect are under consideration and interest in almost all different fields of physics in recent years [2-3]. Although the study of the Casimir effect has been now developed to applications in nanotechnology, there are still research interests in theoretical calculation of the Casimir effect for different geometries. Although the Casimir effect have been already studied for a number of geometries (e.g. parallel plates, long cylinders, spheres, wedges) based on different methods and techniques [4-13], there are a few references in which the geometry of a cylindrical pillbox is considered not for calculating the Casimir force in them as a single cavity but as a surrounding in which the interaction of a partition (plate) with one of its bases is studied [14-15]. Here, we want to study the Casimir effect in a single conducting cylindrical cavity. The significance of studying such a geometry, in addition to its theoretical considerations in physics, lies in its possible application in new world of nanotechnology; this is because this particular geometry is directly related to the finite nanotubes and their applications. Our approach is mathematically based on the Green function method, and conceptually based on virtual photons in the quantum vacuum ocean. We work in a covariant formalism with the quantized electromagnetic four potential $A^{\mu}$ under Dirichlet boundary condition where we have explained about its preferences in [16].

## The Vacuum energy-momentum tensor and the Green Function

Using the symmetric electromagnetic energy-momentum tensor:

$$T^{\alpha\beta} = g^{\alpha\mu}F_{\mu\lambda}F^{\lambda\beta} + \frac{1}{4}g^{\alpha\beta}F_{\mu\upsilon}F^{\mu\upsilon} \qquad (1)$$

where $F^{\mu\nu}(x) = \partial^{\mu}A^{\nu}(x) - \partial^{\nu}A^{\mu}(x)$ is the electromagnetic field tensor and $A^{\mu}(x) = (\phi, \vec{A})$ is the four electromagnetic potential in the Minkowski space-time $(ct, \vec{x})$ and satisfies the wave equation $\partial_{\mu}\partial^{\mu}A^{\upsilon} = 0$ ( $g^{\mu\nu} = diag(1,-1,-1,-1)$ ).
Clearly:

$$T^{\alpha\beta}(x)=\partial^{\alpha}A_{\lambda}\partial^{\lambda}A^{\beta}-\partial^{\alpha}A_{\lambda}\partial^{\beta}A^{\lambda}-\partial_{\lambda}A^{\alpha}\partial^{\lambda}A^{\beta}+\partial_{\lambda}A^{\alpha}\partial^{\beta}A^{\lambda}+\frac{1}{2}g^{\alpha\beta}\left(\partial_{\mu}A_{\upsilon}\partial^{\mu}A^{\upsilon}-\partial_{\mu}A_{\upsilon}\partial^{\upsilon}A^{\mu}\right)$$

$$=\lim_{x\to x'}\Big[\partial^{\alpha}\partial'^{\lambda}A_{\lambda}(x)A^{\beta}(x')-\partial^{\alpha}\partial'^{\beta}A_{\lambda}(x)A^{\lambda}(x')-\partial_{\lambda}\partial'^{\lambda}A^{\alpha}(x)A^{\beta}(x')+\partial_{\lambda}\partial'^{\beta}A^{\alpha}(x)A^{\lambda}(x')$$

$$+1/2g^{\alpha\beta}\left(\partial_{\mu}\partial'^{\mu}A_{\upsilon}(x)A^{\upsilon}(x')-\partial_{\mu}\partial'^{\upsilon}A_{\upsilon}(x)A^{\mu}(x')\right)\Big] \qquad (2)$$

As a general covariant formula, the quantum vacuum expectation value of the energy-momentum tensor operator is found as [16]:

$$\langle 0|\hat{T}^{\alpha\beta}|0\rangle=2i\hbar c\lim_{x\to x'}(\partial^{\alpha}\partial'^{\beta}-\frac{1}{4}g^{\alpha\beta}\partial_{\mu}\partial'^{\mu})G(x,x') \qquad (3)$$

Consider a finite cylindrical box with a circular basis of radius $a$ and length (height) $b$. Among a number of different possible forms of the Green function (under Dirichlet boundary condition), we choose the following form in terms of the $n$ th root of the Bessel functions:

$$G_{in}(x,x')=\frac{1}{\pi^{2}a^{2}b}\sum_{m=-\infty}^{\infty}\sum_{n=1}^{\infty}\sum_{l=1}^{\infty}\int d\omega\frac{e^{-i\omega(t-t')}e^{im(\varphi-\varphi')}}{\lambda_{mnl}(\omega)J_{m+1}^{2}(x_{mn})}J_{m}\left(x_{mn}\frac{\rho}{a}\right)J_{m}\left(x_{mn}\frac{\rho'}{a}\right)$$
$$\times\sin\left(\frac{l\pi z}{b}\right)\sin\left(\frac{l\pi z'}{b}\right) \qquad (4)$$

where $x_{mn}$ is the $n$ th root of the Bessel function of order $m$ and $\lambda_{mnl}(\omega)=-c\left[-\frac{\omega^{2}}{c^{2}}+\frac{x_{mn}^{2}}{a^{2}}+\left(\frac{l\pi}{b}\right)^{2}\right]$.

**The total Casimir energy inside the Cavity**

We want to find the total vacuum energy ( $E_{in}=\int\left\langle\hat{T}^{00}\right\rangle_{in}d^{3}x$ ) inside the cavity. Using (3) and (4):

$$\left\langle\hat{T}^{00}\right\rangle_{in}=-\frac{i\hbar c}{2\pi^{2}a^{2}b}\sum_{m=-\infty}^{\infty}\sum_{n=1}^{\infty}\sum_{l=1}^{\infty}\int\frac{d\omega}{c}\frac{1}{J_{m+1}^{2}(x_{mn})}\frac{1}{-\frac{\omega^{2}}{c^{2}}+\frac{x_{mn}^{2}}{a^{2}}+\left(\frac{l\pi}{b}\right)^{2}}$$
$$\times\left\{3\frac{\omega^{2}}{c^{2}}J_{m}^{2}\left(x_{mn}\frac{\rho}{a}\right)\sin^{2}\left(\frac{l\pi z}{b}\right)+\left[\left(\frac{x_{mn}}{a}\right)^{2}\left(J'_{m}\left(x_{mn}\frac{\rho}{a}\right)\right)^{2}+\frac{m^{2}}{\rho^{2}}J_{m}^{2}\left(x_{mn}\frac{\rho}{a}\right)\right]\sin^{2}\left(\frac{l\pi z}{b}\right)\right.$$
$$\left.+\left(\frac{l\pi}{b}\right)^{2}J_{m}^{2}\left(x_{mn}\frac{\rho}{a}\right)\cos^{2}\left(\frac{l\pi z}{b}\right)\right\} \qquad (5)$$

where $J'_{m}\left(x_{mn}\frac{\rho}{a}\right)=\frac{a}{x_{mn}}\frac{d}{d\rho}J_{m}\left(x_{mn}\frac{\rho}{a}\right)$.

The physical result is found by the application of complex frequency rotation $\omega \rightarrow i\omega$:

$$\left\langle \hat{T}^{00} \right\rangle_{in} = \frac{\hbar c}{2\pi^2 a^2 b} \sum_{m=-\infty}^{\infty} \sum_{n=1}^{\infty} \sum_{l=1}^{\infty} \int \frac{d\omega}{c} \frac{1}{J_{m+1}^2(x_{mn})} \frac{1}{\frac{\omega^2}{c^2} + \frac{x_{mn}^2}{a^2} + \left(\frac{l\pi}{b}\right)^2}$$

$$\times \left\{ -3\frac{\omega^2}{c^2} J_m^2\left(x_{mn}\frac{\rho}{a}\right) \sin^2\left(\frac{l\pi z}{b}\right) + \left[ \left(\frac{x_{mn}}{a}\right)^2 \left(J_m'\left(x_{mn}\frac{\rho}{a}\right)\right)^2 + \frac{m^2}{\rho^2} J_m^2\left(x_{mn}\frac{\rho}{a}\right) \right] \sin^2\left(\frac{l\pi z}{b}\right) + \left(\frac{l\pi}{b}\right)^2 J_m^2\left(x_{mn}\frac{\rho}{a}\right) \cos^2\left(\frac{l\pi z}{b}\right) \right\} \quad (6)$$

Knowing:

$$\int_0^a \left[ \left(\frac{x_{mn}}{a}\right)^2 \left(J_m'\left(x_{mn}\frac{\rho}{a}\right)\right)^2 + \frac{m^2}{\rho^2} J_m^2\left(x_{mn}\frac{\rho}{a}\right) \right] \rho\, d\rho = \frac{x_{mn}^2}{2} J_{m+1}^2(x_{mn}) \qquad m \geq 0 \qquad (7),$$

the total internal vacuum energy is found as:

$$E_{in} = \int_0^a \int_0^{2\pi} \int_0^b \left\langle \hat{T}^{00} \right\rangle_{in} \rho\, d\rho\, d\varphi\, dz = \frac{\hbar c}{4\pi} \sum_{m=0}^{\infty} \eta_m \sum_{n=1}^{\infty} \sum_{l=1}^{\infty} \int_0^{\infty} \frac{d\omega}{c} \frac{-3\frac{\omega^2}{c^2} + \frac{x_{mn}^2}{a^2} + \left(\frac{l\pi}{b}\right)^2}{\frac{\omega^2}{c^2} + \frac{x_{mn}^2}{a^2} + \left(\frac{l\pi}{b}\right)^2} \qquad (8)$$

where $\eta_m = \begin{cases} 1 & m = 0 \\ 2 & m \neq 0 \end{cases}$.

**Repulsive Casimir force in the cavity**

Assuming the walls of the cavity under consideration are made of good conducting materials (metals), we can apply the plasma frequency cutoff regularization in (8) [16]. For a cavity of a size at the scales corresponding to the current micro(nano)scopic world ($a \sim b \sim 10^{-7} m$), knowing that the plasma frequency of good real conductors (e.g. gold, silver, copper) is at the order of $\omega_p \leq 10^{16} rad/\sec$ [17], and that the roots of the Bessel functions start from 2.40482 to greater values, it is clear that $\frac{\omega_p^2}{c^2} << \frac{x_{mn}^2}{a^2} + \left(\frac{l\pi}{b}\right)^2$ for enough large values of $m = M, n = N, l = N$; so, one can simply divide the right hand side of (8) into two parts as in the following:

$$E_{in}=\frac{\hbar c}{4\pi}\left\{\sum_{m=0}^{M}\eta_m\sum_{n=1}^{N}\sum_{l=1}^{L}\int_0^{\omega_p}\frac{d\omega}{c}\frac{-3\frac{\omega^2}{c^2}+\frac{x_{mn}^2}{a^2}+\left(\frac{l\pi}{b}\right)^2}{\frac{\omega^2}{c^2}+\frac{x_{mn}^2}{a^2}+\left(\frac{l\pi}{b}\right)^2}+\sum_{m=M+1}^{\infty}\eta_m\sum_{n=N+1}^{\infty}\sum_{l=L+1}^{\infty}\int_0^{\omega_p}\frac{d\omega}{c}\right\} \quad (9)$$

where $M$, $N$, and $L$ are finite constants of large values. The second part in (9) is an infinite constant that does not depend on $a$ and $b$ and thus has no contribution to the Casimir force which is derived as ($-\nabla_{a,b}E$). In other words, although we deal with an infinite value for the Casimir energy, the physical values of the Casimir forces under consideration are finite because they are finite polynomials and not infinite series.

Integrating out (9) leads to:

$$E_{in}=\frac{\hbar c}{\pi}\sum_{m=0}^{M}\eta_m\sum_{n=1}^{N}\sum_{l=1}^{L}\sqrt{\frac{x_{mn}^2}{a^2}+\left(\frac{l\pi}{b}\right)^2}\arctan\left[\frac{\frac{\omega_p}{c}}{\sqrt{\frac{x_{mn}^2}{a^2}+\left(\frac{l\pi}{b}\right)^2}}\right]+\text{infinite constant} \quad (10)$$

To calculate the desired Casimir forces, we should find the outer space energy $E_{out}$ too, but it can be explained that the external energy has a negligible contribution [1] and thus it is enough to compute the following local Casimir forces:

$$F_a=-\frac{\partial E_{in}}{\partial a}=\frac{\hbar c}{\pi a^3}\sum_{m=0}^{M}\eta_m\sum_{n=1}^{N}\sum_{l=1}^{L}x_{mn}^2\left\{\frac{1}{\sqrt{\frac{x_{mn}^2}{a^2}+\left(\frac{l\pi}{b}\right)^2}}\arctan\left[\frac{\frac{\omega_p}{c}}{\sqrt{\frac{x_{mn}^2}{a^2}+\left(\frac{l\pi}{b}\right)^2}}\right]-\frac{\frac{\omega_p}{c}}{\frac{\omega_p^2}{c^2}+\frac{x_{mn}^2}{a^2}+\left(\frac{l\pi}{b}\right)^2}\right\} \quad (11)$$

$$F_b = -\frac{\partial E_{in}}{\partial b} = \frac{\pi\hbar c}{b^3}\sum_{m=0}^{M}\eta_m \sum_{n=1}^{N}\sum_{l=1}^{L} l^2 \left\{ \frac{1}{\sqrt{\frac{x_{mn}^2}{a^2}+\left(\frac{l\pi}{b}\right)^2}} \arctan\left[\frac{\frac{\omega_p}{c}}{\sqrt{\frac{x_{mn}^2}{a^2}+\left(\frac{l\pi}{b}\right)^2}}\right] - \frac{\frac{\omega_p}{c}}{\frac{\omega_p^2}{c^2}+\frac{x_{mn}^2}{a^2}+\left(\frac{l\pi}{b}\right)^2} \right\} \tag{12}$$

It is clear that the Casimir forces experienced by the bases and the lateral surface are repulsive.

Introducing $y_p = \frac{\omega_p}{c}a$, and $a = \alpha b$ ($\alpha$ is of the order of 1):

$$F_a = \frac{\hbar c}{\pi a^2}\sum_{m=0}^{M}\eta_m \sum_{n=1}^{N}\sum_{l=1}^{L} x_{mn}^2 \left\{ \frac{1}{\sqrt{x_{mn}^2+(l\pi\alpha)^2}} \arctan\left[\frac{y_p}{\sqrt{x_{mn}^2+(l\pi\alpha)^2}}\right] - \frac{y_p}{y_p^2+x_{mn}^2+(l\pi\alpha)^2} \right\} \tag{13}$$

$$F_b = \frac{\hbar c\pi\alpha^3}{a^2}\sum_{m=0}^{M}\eta_m \sum_{n=1}^{N}\sum_{l=1}^{L} l^2 \left\{ \frac{1}{\sqrt{x_{mn}^2+(l\pi\alpha)^2}} \arctan\left[\frac{y_p}{\sqrt{x_{mn}^2+(l\pi\alpha)^2}}\right] - \frac{y_p}{y_p^2+x_{mn}^2+(l\pi\alpha)^2} \right\} \tag{14}$$

Using Wolfram Mathematica 7 algorithm for the computation of the roots of Bessel's functions, and Borland Delphi 7 programming for computing the series (sums), with $\omega_p \sim 10^{16}\, rad/\sec$, $a = 10^{-7}m$, $\alpha = 1$, $F_a(F_b)$ values up to the $M = 500$, $N = 500$, $L = 500$ compared with when $M = 499$, $N = 499$, $L = 499$ are computed as:

$$\begin{aligned} F_a\big|_{M=N=L=500} &= 3.746097(nanoNewton) \\ F_b\big|_{M=N=L=500} &= 2.108897(nanoNewton) \\ F_a\big|_{M=N=L=499} &= 3.738501(nanoNewton) \\ F_b\big|_{M=N=L=499} &= 2.104639(nanoNewton) \end{aligned} \tag{15}$$

Thus:

$$\frac{F_a(500)}{F_a(499)} = \frac{F_b(500)}{F_b(499)} \cong 1.002 \qquad (16)$$

This shows if we keep $500^3 = 125$ millions terms, we shall have an acceptable result with a relative error of $0.002$. More details of our numerical computations and a number of data even for $\alpha = \frac{a}{b} \neq 1$ are in the tables in the Appendix.

**Asymptotic behavior**

We expect to find out the result for a long cylinder of circular basis of radius $a$ in the limit of $b >> a$. With the change of variable $\frac{1}{b}\sum \rightarrow \sim \int dk$, the Casimir energy per unit height of the long cylinder is found as:

$$\frac{E}{b} \sim \sum_{m,n} \int dk \frac{d\omega}{c} \frac{-3\frac{\omega^2}{c^2} + \frac{x_{mn}^2}{a^2} + k^2}{\frac{\omega^2}{c^2} + \frac{x_{mn}^2}{a^2} + k^2} \qquad (17)$$

Application of $\int dk \frac{d\omega}{c} \rightarrow \int \lambda d\lambda d\theta$ with $\lambda^2 = \frac{\omega^2}{c^2} + k^2$ results in:

$$\frac{E}{b} \sim \frac{1}{a^2} \sum_{m,n} x_{mn}^2 \ln\left(1 + \frac{\lambda_p^2 a^2}{x_{mn}^2}\right) + \text{a term independent of } a \qquad (18)$$

The corresponding Casimir force ($\frac{F}{b} \sim -\frac{\partial}{\partial a}(\frac{E}{b})$) is proportional to the cubic inverse of the radius ($\frac{F}{b} \propto \frac{1}{a^3}$); this is an expectable result [16, 18].

As another asymptotic behavior, let check if we can arrive at the well-known result of the Casimir force (pressure) for two parallel conducting plates in the limit of $a >> b$. The relation (8) can be simply written in the following form:

$$E = \frac{\hbar c}{8\pi} \sum_{n=0}^{\infty} \eta_m \sum_{n=1}^{\infty} \int \frac{d\omega}{c} \times \left\{ \left( -3\frac{\omega^2}{c^2} + \frac{x_{mn}^2}{a^2} \right) \left( \frac{b}{\beta_{mn}} \coth(\beta_{mn} b) - \frac{1}{\beta_{mn}^2} \right) - (\beta_{mn} b)\coth(\beta_{mn} b) - 1 \right\} \tag{19}$$

where $\beta_{mn}^2 = \frac{\omega^2}{c^2} - \frac{x_{mn}^2}{a^2}$.

In the continuous limit $\frac{x_{mn}}{a} \to k$, $\frac{1}{\pi a^2} \sum_m \sum_n \sim \int_0^{\infty} k dk$, and $\beta_{mn}^2 \to \beta^2 = \frac{\omega^2}{c^2} + k^2$, it is found that:

$$\frac{E}{\pi a^2} \sim \int k dk \frac{d\omega}{c} \left\{ \left( k^2 - 3\frac{\omega^2}{c^2} \right) \left( \frac{b}{\beta} \coth(\beta b) - \frac{1}{\beta^2} \right) - (\beta b)\coth(\beta b) - 1 \right\} \tag{20}$$

Or:

$$\frac{E}{area} \sim -\int k dk \frac{d\omega}{c} \left\{ 4\frac{\omega^2}{c^2} \frac{b}{\beta} \coth(\beta b) \right\} + \text{constant} \tag{21}$$

Using $\frac{\omega}{c} = \beta \cos\theta$, and $\int_0^{\infty} k dk \int_{-\infty}^{\infty} \frac{d\omega}{c} = \int_{-1}^{1} d\cos\theta \int_0^{\infty} \beta^2 d\beta$:

$$\frac{E}{area} \sim -b \left( \int_0^{\infty} \beta^3 d\beta \frac{e^{2\beta b} + 1}{e^{2\beta b} - 1} \right) = -\frac{I(p)}{b^3} \tag{22}$$

where $I(p) = \int_0^{p} s^3 ds \frac{e^s + 1}{e^s - 1}$ is a cutoff number whose value depends on the cutoff frequency ( $p = 2\beta_p b$ ). Thus, the Casimir force per unit area (pressure) is

$\left(\frac{F}{area} \propto -\frac{1}{b^4}\right)$.

## Conclusion

In conclusion, we have solved the problem of the Casimir effect for a single conducting microscopic cylindrical cavity. The quantized field under consideration is the electromagnetic four potential $A^{\mu}$ with the advantages mentioned in [16]; of course, one can simply repeat all the calculation for scalar fields with only a numerical coefficient difference in the result.

A number of mathematical, physical, and fundamental tools, including choosing an appropriate Green function, applying plasma frequency cutoff regularization, using the uncertainty relations' limits on virtual particles [1], have been used to find out the resulting Casimir forces. It is natural to find out a repulsive result not only for the cylindrical cavity we have studied here but also for any other closed (topologically compact) object. This can be simply justified by considering the origin of the Casimir effect based on the virtual particles of the vacuum "sea"; when the "free" virtual photons are confined in a smaller closed volume by the presence of a boundary (e.g. cylindrical pillbox), they push out the surfaces more stronger than the outer virtual photons which moved more freely in the outer space. A repulsive Casimir force is interesting in MEMS (NEMS) manufacturing for the omission of the stiction due to the attracting Casimir effect (e.g. [19]). As was seen, the asymptotic behavior is expectable and may be considered as a supporting tool for why we have particularly worked with $\alpha = \frac{a}{b} \approx 1$.

## Appendix

Introducing the new dimensionless variables:

$$I_a = \sum_{m=0}^{M} \eta_m \sum_{n=1}^{N} \sum_{l=1}^{L} x_{mn}^2 \left\{ \frac{1}{\sqrt{x_{mn}^2 + (l\pi\alpha)^2}} \arctan\left[ \frac{y_p}{\sqrt{x_{mn}^2 + (l\pi\alpha)^2}} \right] - \frac{y_p}{y_p^2 + x_{mn}^2 + (l\pi\alpha)^2} \right\} \tag{A-1}$$

$$I_b = \pi^2 \alpha^3 \sum_{m=0}^{M} \eta_m \sum_{n=1}^{N} \sum_{l=1}^{L} l^2 \left\{ \frac{1}{\sqrt{x_{mn}^2 + (l\pi\alpha)^2}} \arctan\left[ \frac{y_p}{\sqrt{x_{mn}^2 + (l\pi\alpha)^2}} \right] - \frac{y_p}{y_p^2 + x_{mn}^2 + (l\pi\alpha)^2} \right\} \tag{A-2}$$

**Table 1**: Details of the numerical computation of the Casimir forces for $\alpha = 1$

| $M, N, L$ | $I_a$ | $I_b$ |
|---|---|---|
| 10 | 51.18525157 | 32.03410391 |
| 50 | 338.88613822 | 196.32843402 |
| 100 | 710.77177245 | 405.98440388 |
| 150 | 1084.87833894 | 616.37200613 |
| 200 | 1460.99719072 | 827.48128124 |
| 250 | 1837.25695257 | 1038.61541021 |
| 300 | 2213.89567974 | 1249.87660125 |
| 350 | 2590.79582075 | 1461.22535291 |
| 400 | 2967.88724331 | 1672.63812911 |
| 450 | 3345.12471036 | 1884.09976349 |
| 499 | 3714.92937805 | 2091.36960004 |
| 500 | 3722.47740714 | 2095.59992939 |

**Table 2**: Details of the numerical computation of the Casimir forces for $\alpha = 1.5$

| $M, N, L$ | $I_a$ | $I_b$ |
|---|---|---|
| 10 | 32.88324866 | 41.12744619 |
| 50 | 229.16927677 | 242.99717456 |
| 100 | 485.12930880 | 499.57391866 |
| 150 | 743.14419686 | 756.77925636 |
| 200 | 1002.73136023 | 1014.90323416 |
| 250 | 1262.54638094 | 1272.95265349 |
| 300 | 1522.69700396 | 1531.12892691 |
| 350 | 1783.07927549 | 1789.39264178 |
| 400 | 2043.63112467 | 2047.72030657 |
| 450 | 2304.31249412 | 2306.09677583 |
| 499 | 2559.87953545 | 2559.34312439 |
| 500 | 2565.09607223 | 2564.51174954 |

**Table 3**: Details of the numerical computation of the Casimir forces for $\alpha = 0.25$

| $M, N, L$ | $I_a$ | $I_b$ |
|---|---|---|
| 10 | 136.17949081 | 6.63129134 |
| 50 | 918.31704267 | 59.73948949 |
| 100 | 1925.15532407 | 131.01361910 |
| 150 | 2936.86472139 | 203.24927232 |
| 200 | 3952.62050083 | 275.88540881 |
| 250 | 4968.73690218 | 348.72149449 |
| 300 | 5985.63132119 | 421.68792299 |
| 350 | 7003.06007385 | 494.74379749 |
| 400 | 8020.87859285 | 567.86487951 |
| 450 | 9038.99401519 | 641.03561551 |
| 499 | 10036.9742918 | 712.78090146 |
| 500 | 10057.3432534 | 714.24542821 |

**Table 4**: Details of the numerical computation of the Casimir forces for $\alpha = 4$

| $M, N, L$ | $I_a$ | $I_b$ |
|---|---|---|
| 10 | 8.17224682 | 53.96587042 |
| 50 | 77.92634249 | 322.25927118 |
| 100 | 173.66631302 | 661.84610177 |
| 150 | 271.26095744 | 1001.93722861 |
| 200 | 369.86672978 | 1343.18941429 |
| 250 | 468.81498484 | 1684.24520411 |
| 300 | 568.04472194 | 2025.42770747 |
| 350 | 667.46894868 | 2366.69757181 |
| 400 | 767.03565315 | 2708.03133569 |
| 450 | 866.71124350 | 3049.41386448 |
| 499 | 964.47681957 | 3384.00614021 |
| 500 | 966.47277819 | 3390.83488602 |